\documentclass[a4paper]{spie}
\usepackage{graphicx}
\usepackage{amsfonts,amssymb,amsbsy,amsmath}

\usepackage{color}
\usepackage[normalem]{ulem}

\title{Measuring and engineering the atomic mass density wave\\ of a Gaussian mass-polariton pulse in optical fibers} 
\author{Mikko Partanen and Jukka Tulkki\skiplinehalf
Engineered Nanosystems group, School of Science\\Aalto University, P.O. Box 12200, 00076 Aalto, Finland\\
}

\begin{document} 

\maketitle 

\begin{abstract}
Conventional theories of electromagnetic waves in a medium assume that only the
energy of the field propagates inside the medium. Consequently, they neglect the
transport of mass density by the medium atoms. We have recently presented foundations
of a covariant theory of light propagation in a nondispersive medium by considering
a light wave simultaneously with the dynamics of the medium atoms driven by optoelastic
forces [Phys.~Rev.~A \textbf{95}, 063850 (2017)].
In particular, we have shown that the mass is transferred by an atomic mass density wave (MDW),
which gives rise to mass-polariton (MP) quasiparticles, i.e., covariant coupled
states of the field and matter having a nonzero rest mass. Another key
observation of the mass-polariton theory of light is that, in common semiconductors,
most of the momentum of light is transferred by moving atoms, e.g., 92\% in the case of silicon.
In this work, we generalize the MP theory of light for dispersive media and
consider experimental measurement of the mass transferred by the MDW
atoms when an intense light pulse propagates in a silicon fiber.
In particular, we consider optimal intensity and time dependence of a Gaussian pulse
and account for the breakdown threshold irradiance of the material.
The optical shock wave property of the MDW, which propagates
with the velocity of light instead of the velocity of sound, prompts for
engineering of novel device concepts like very high frequency mechanical
oscillators not limited by the acoustic cutoff frequency.
\end{abstract}

\section{Introduction}

It is well known in the electrodynamics of continuous media that a light pulse
exerts optical forces on the atoms when it propagates inside a medium \cite{Landau1984}.
However, the coupled dynamics of the field and the medium driven by the optical
force density has been a subject of very few detailed studies. We have recently elaborated how
the optical force density gives rise to a mass density wave (MDW) in a nondispersive medium when
a light pulse is propagating in it \cite{Partanen2017c}. We have also generalized the
theory for dispersive media \cite{Partanen2017e}. Our investigations
are performed using two complementary approaches:
(1) the mass-polariton (MP) quasiparticle theory, which is based on the conservation laws
and the Lorentz transformation; (2) the classical optoelastic continuum dynamics (OCD), which
generalizes the electrodynamics of continuous media to include the dynamics of the medium
resulting from the optical force density.

The complementary MP and OCD models transparently resolve
the centenary Abraham-Minkowski controversy of the momentum of light in a medium
\cite{Leonhardt2006,Cho2010,Pfeifer2007,Barnett2010b,Barnett2010a,Leonhardt2014,Saldanha2017,Ramos2015,Brevik2017,Mansuripur2010,Kemp2011},
which has also gained much experimental interest
\cite{Choi2017,Campbell2005,Sapiro2009,Jones1954,Jones1978,Walker1975,She2008,Zhang2015,Astrath2014,Ashkin1973,Casner2001,Brevik1979}.
This controversy originates from the formulation of two rivaling
momentum densities for light by Abraham,
$\mathbf{g}_\mathrm{A}=\mathbf{E}\times\mathbf{H}/c^2$
\cite{Abraham1909,Abraham1910}, and by Minkowski,
$\mathbf{g}_\mathrm{M}=\mathbf{D}\times\mathbf{B}$ \cite{Minkowski1908},
where $\mathbf{E}$ and $\mathbf{H}$
are the electric- and magnetic-field strengths, $\mathbf{D}$ and
$\mathbf{B}$ are the electric and magnetic
flux densities, and $c$ is the speed of light in vacuum.
In contrast, in the MP theory of light, the total momentum density of light in the medium is given by
$\mathbf{g}_\mathrm{MP}=\rho_\mathrm{a}\mathbf{v}_\mathrm{a}+\mathbf{E}\times\mathbf{H}/c^2$,
where $\rho_\mathrm{a}\mathbf{v}_\mathrm{a}$ is the classical momentum
density of the MDW atoms. For a monochromatic field, the momentum
densities $\mathbf{g}_\mathrm{A}$, $\mathbf{g}_\mathrm{M}$, and $\mathbf{g}_\mathrm{MP}$
correspond to the single-photon momenta
$p_\mathrm{A}=\hbar\omega/(n_\mathrm{g}c)$, $p_\mathrm{M}=n_\mathrm{p}^2\hbar\omega/(n_\mathrm{g}c)$,
and $p_\mathrm{MP}=n_\mathrm{p}\hbar\omega/c$,
respectively, where $\hbar$ is the reduced Planck constant, $\omega$ is the angular
frequency of the field, and $n_\mathrm{p}$ and $n_\mathrm{g}$ are the phase
and group refractive indices of the medium.

In this work, we review the foundations of the MP theory of light
represented in Refs.~\cite{Partanen2017c,Partanen2017e,Partanen2016a,Partanen2017d},
especially concentrating in the OCD model and the generalization of the theory for dispersive media.
We also consider experimental measurement of the mass transferred by the MDW atoms
when an intense light pulse propagates in a silicon fiber. In particular, we study optimal intensity and time
dependence of a Gaussian pulse and account for the breakdown threshold irradiance of the material.

\section{Fields and the dispersion relation}
\label{sec:fieldsolution}

In dispersive media, the phase refractive index $n_\mathrm{p}(\omega)$ depends on frequency.
Its general relation to the wave number $k$ and frequency $\omega(k)$ is given by $\omega(k)=ck/n_\mathrm{p}(\omega)$,
which can be used to write the phase velocity of light in a medium as
$v_\mathrm{p}(\omega)=c/n_\mathrm{p}(\omega)=\omega(k)/k$.
The phase velocity describes the propagation velocity of individual
frequency components. Another velocity that characterizes the propagation
of light in dispersive media is provided by the group
velocity $v_\mathrm{g}(\omega)=c/n_\mathrm{g}(\omega)=\partial\omega(k)/\partial k$,
where $n_\mathrm{g}(\omega)$ is the group refractive index.
The group velocity describes the amplitudes of individual frequency components
together produce a wave packet.

As exact solutions of Maxwell's equations,
the electric and magnetic fields of a one-dimensional
light pulse that is linearly polarized along the $y$ axis
and propagates in $x$ direction in a dispersive
medium are written as \cite{Griffiths1998}
\begin{equation}
 \mathbf{E}(\mathbf{r},t)
 =\mathrm{Re}\Big[\int_{-\infty}^\infty \tilde E(k)e^{i[kx-\omega(k) t]}dk\Big]\hat{\mathbf{y}},
 \label{eq:electricfieldgeneral}
\end{equation}
\begin{equation}
 \mathbf{H}(\mathbf{r},t)
 =\mathrm{Re}\Big[\int_{-\infty}^\infty \tilde H(k)e^{i[kx-\omega(k) t]}dk\Big]\hat{\mathbf{z}}.
 \label{eq:magneticfieldgeneral}
\end{equation}
Here $\tilde E(k)$ and $\tilde H(k)$ are the Fourier components of the
electric and magnetic fields, and $\hat{\mathbf{y}}$ and $\hat{\mathbf{z}}$
are unit vectors with respect to $y$ and $z$ axes, respectively. The Fourier components
of the fields are related to each other by
$\tilde H(k)=\sqrt{\varepsilon[\omega(k)]/\mu[\omega(k)]}\tilde E(k)$,
where $\varepsilon(\omega)$ and $\mu(\omega)$ are the
permittivity and permeability of the medium, which are frequency dependent in dispersive media.
The permittivity and permeability are related to the phase refractive index of the medium by
$\varepsilon(\omega)\mu(\omega)=\varepsilon_0\mu_0n_\mathrm{p}(\omega)^2$,
in which $\varepsilon_0$ and $\mu_0$ are the permittivity and permeability
of the vacuum, respectively.

In this work, we concentrate on the propagation of light pulses in
the special case of linearly dispersive media. Linear dispersion is a good approximation
for light pulses with the central frequency $\omega_0$ in any dispersive media if
the spectral width $\Delta \omega/\omega_0$ of the wave packet is relatively small,
the dispersion relation does not have resonances near $\omega_0$,
and if the distance traveled by the wave packet is not very long.
The linear dispersion relation near the central frequency $\omega_0$
can be written as
\begin{equation}
\omega(k)=\omega_0+(c/n_\mathrm{g})(k-k_\mathrm{0,med}),
\label{eq:lineardispersion}
\end{equation}
where $k_\mathrm{0,med}=n_\mathrm{p}k_0$ is the central wave number
in the medium, in which $k_0=\omega_0/c$ is the central wave number in vacuum and
$n_\mathrm{p}=n_\mathrm{p}(\omega_0)$ is the phase refractive index for $\omega_0$,
and the group refractive index $n_\mathrm{g}$ is constant.
The special feature of the linear form of the dispersion relation in Eq.~\eqref{eq:lineardispersion} is
that it does not lead to the distortion of the pulse envelope while the pulse propagates.
For frequencies deviating from $\omega_0$,
the linear dispersion relation in Eq.~\eqref{eq:lineardispersion}
defines the frequency-dependent phase refractive index as
$n_\mathrm{p}(\omega)=n_\mathrm{g}+(n_\mathrm{p}-n_\mathrm{g})\omega_0/\omega$.

In this work, we make simulations using Gaussian light pulses with
$\tilde E(k)=\tilde E_0e^{-[(k-n_\mathrm{p}k_0)/(n_\mathrm{p}\Delta k_0)]^2/2}$,
where $\tilde E_0$ is a normalization factor
and $\Delta k_0$ is the standard deviation
of the wave number in vacuum, which determines the pulse width
in the $x$ direction as $\Delta x=1/(\sqrt{2}n_\mathrm{p}\Delta k_0)$.
The standard deviation of the pulse width in time is given by
$\Delta t=n_\mathrm{p}\Delta x/c=1/(\sqrt{2}\Delta k_0c)$
and the corresponding full width at half maximum (FWHM) is $\Delta t_\mathrm{FWHM}=2\sqrt{2\ln 2}\,\Delta t$.
Applying the linear dispersion relation in Eq.~\eqref{eq:lineardispersion}
with the expression of the electric field in Eq.~\eqref{eq:electricfieldgeneral}, 
the integration with respect to $k$ can be performed analytically and the electric field becomes
\begin{equation}
 \mathbf{E}(\mathbf{r},t)
 =\sqrt{2\pi}\,n_\mathrm{p}\Delta k_0\tilde E_0\cos\Big[n_\mathrm{p}k_0\Big(x-\frac{ct}{n_\mathrm{p}}\Big)\Big]
 e^{-(n_\mathrm{p}\Delta k_0)^2(x-ct/n_\mathrm{g})^2/2}\hat{\mathbf{y}}.
 \label{eq:electricfield}
\end{equation}
In the simulations, we determine the normalization factor $\tilde E_0$ in
Eq.~\eqref{eq:electricfield} so that the integral of the corresponding instantaneous
energy density over $x$ gives $U_0/A$,
where $U_0$ is the total electromagnetic energy of the light pulse
and $A$ is the cross-sectional area.

\section{Optoelastic continuum dynamics}
\label{sec:ocdmodel}

\subsection{Newton's equation of motion and optoelastic forces}

In the OCD model, a light pulse is optoelastically coupled to 
the medium through the optical force density,
which gives rise to an atomic MDW in the medium. Due to the MDW effect,
the mass density of the medium $\rho_\mathrm{a}$ is perturbed from its equilibrium value $\rho_0$.
For a light pulse propagating in a homogeneous medium, the perturbed
mass density can be written as
$\rho_\mathrm{a}(\mathbf{r},t)=\rho_0+\rho_\text{\tiny MDW}(\mathbf{r},t)$,
where $\rho_\text{\tiny MDW}(\mathbf{r},t)$ is the mass density of the MDW.
As the atomic velocities are nonrelativistic, we can write
Newton's equation of motion for the mass density
of the medium $\rho_\mathrm{a}(\mathbf{r},t)$ as \cite{Partanen2017c}
\begin{equation}
 \rho_\mathrm{a}(\mathbf{r},t)\frac{d^2\mathbf{r}_\mathrm{a}(\mathbf{r},t)}{dt^2}=\mathbf{f}_\mathrm{opt}(\mathbf{r},t)+\mathbf{f}_\mathrm{el}(\mathbf{r},t).
 \label{eq:mediumnewton}
\end{equation}
Here $\mathbf{r}_\mathrm{a}(\mathbf{r},t)$ is the atomic
displacement field of the medium, $\mathbf{f}_\mathrm{opt}(\mathbf{r},t)$
is the optical force density that will be discussed in more detail below, and
$\mathbf{f}_\mathrm{el}(\mathbf{r},t)$ is the elastic force
density that acts between atoms as they are displaced from
their initial equilibrium positions by the optical force density.
The elastic force density for anisotropic cubic
crystals, such as silicon, is given, e.g., in Ref.~\cite{Kittel2005}.

The correct form of the optical force density acting on the medium under the
influence of time-dependent electromagnetic field has
been extensively discussed in previous literature as a part of the Abraham-Minkowski controversy \cite{Milonni2010}.
In the case of a nondispersive medium, we have recently shown that there is only
one form of optical force density that is fully consistent
with the MP quasiparticle model and the underlying principles
of the special theory of relativity \cite{Partanen2017c}.
In this work, we generalize this optical force for dispersive dielectric media by writing
\begin{equation}
 \mathbf{f}_\mathrm{opt}(\mathbf{r},t)=-\varepsilon_0n_\mathrm{g}\mathbf{E}^2\nabla n_\mathrm{p}+\frac{n_\mathrm{p}n_\mathrm{g}-1}{c^2}\frac{\partial}{\partial t}\mathbf{E}\times\mathbf{H},
 \label{eq:opticalforcedensity}
\end{equation}
where $\mathbf{E}\times\mathbf{H}=\mathbf{S}$ is the instantaneous Poynting vector.
This expression of the optical force density can be justified by showing that
it is the only form of the optical force density that
enables covariant description of the light pulse in a dispersive medium.
In calculating the optoelastic force field by using Eq.~\eqref{eq:opticalforcedensity},
we apply a perturbative approach in which the damping of the electromagnetic field due to
the transfer of the field energy to the kinetic and elastic energies
of atoms by the optical force is neglected. The accuracy of this
approximation is estimated in the case of nondispersive media in Ref.~\cite{Partanen2017c}.
The conclusions are also valid for lossless dispersive media
where the direct optical absorption related, e.g.,
to the electronic excitation of the medium, is negligible.

\subsection{Transferred mass and momentum of the mass density wave}

The optical and elastic force densities
and Newton's equation of motion in Eq.~\eqref{eq:mediumnewton}
can be used to simulate the motion of atoms in the medium
as a function of space and time. The total displacement of atoms at position $\mathbf{r}$
is solved from Eq.~\eqref{eq:mediumnewton} by integration as
\begin{equation}
 \mathbf{r}_\mathrm{a}(\mathbf{r},t)=\int_{-\infty}^t\int_{-\infty}^{t''}\frac{d^2\boldsymbol{r}_\mathrm{a}(\mathbf{r},t')}{dt'^2}dt'dt''
 =\int_{-\infty}^t\int_{-\infty}^{t''}\frac{\mathbf{f}_\mathrm{opt}(\mathbf{r},t')+\mathbf{f}_\mathrm{el}(\mathbf{r},t')}{\rho_\mathrm{a}(\mathbf{r},t')}dt'dt''. 
 \label{eq:displacement}
\end{equation}
The perturbed mass density of the medium $\rho_\mathrm{a}(\mathbf{r},t)$
is extremely close to the equilibrium mass density $\rho_0$ as
the total mass of atoms inside the light pulse is very large
compared to mass equivalent of the field energy. Therefore,
the mass density in the denominator of the integrand in Eq.~\eqref{eq:displacement}
can be approximated with the equilibrium mass density $\rho_0$.

When the light pulse has passed the position $\mathbf{r}$
at the instance of time $t_\mathrm{pass}$, the displacement of atoms
at $\mathbf{r}$ is given by $\mathbf{r}_\mathrm{a}(\mathbf{r},t_\mathrm{pass})$.
One can then obtain the total displaced volume as
$\delta V=\int\mathbf{r}_\mathrm{a}(\mathbf{r},t_\mathrm{pass})\cdot d\mathbf{A}$,
where the integration is performed over the transverse plane
with a surface element vector $d\mathbf{A}$. Using the solution of the atomic displacements
in Eq.~\eqref{eq:displacement}, we can then obtain an equation for the total transferred mass
$\delta M=\rho_0\delta V$ as
\begin{equation}
 \delta M=\int\int_{-\infty}^\infty\int_{-\infty}^{t}[\mathbf{f}_\mathrm{opt}(\mathbf{r},t')+\mathbf{f}_\mathrm{el}(\mathbf{r},t')]dt'dt\cdot d\mathbf{A}.
 \label{eq:mdwmass2}
\end{equation}
The total transferred mass $\delta M$ can also be given in terms of the MDW mass density as
$\delta M=\int\rho_\text{\tiny MDW}(\mathbf{r},t)dV$. Using Eq.~\eqref{eq:mdwmass2} and the relation $cdt=n_\mathrm{g}dx$,
one then obtains the mass density of the MDW, given by
\begin{equation}
 \rho_\text{\tiny MDW}(\mathbf{r},t)=\frac{n_\mathrm{g}}{c}\int_{-\infty}^{t}[\mathbf{f}_\mathrm{opt}(\mathbf{r},t')+\mathbf{f}_\mathrm{el}(\mathbf{r},t')]\cdot\hat{\mathbf{x}}\,dt'.
 \label{eq:mdwdensity}
\end{equation}
Here $\hat{\mathbf{x}}$ is the unit vector in the $x$ direction, i.e., in the direction of light propagation.
Using Eq.~\eqref{eq:mdwdensity} and the optical and elastic force densities, we can perform
numerical simulations of the MDW driven by a light pulse propagating in a dispersive medium.

The velocity field of the medium is given by a time derivative of the atomic displacement field as
$\mathbf{v}_\mathrm{a}(\mathbf{r},t)=d\boldsymbol{r}_\mathrm{a}(\mathbf{r},t)/dt$.
Using the classical momentum density of the medium, given by $\rho_\mathrm{a}\mathbf{v}_\mathrm{a}(\mathbf{r},t)$,
the momentum of the MDW can be directly obtained by integration as
\begin{equation}
 \mathbf{p}_\text{\tiny MDW}=\int \rho_\mathrm{a}\mathbf{v}_\mathrm{a}(\mathbf{r},t)d^3r=\int \rho_\text{\tiny MDW}(\mathbf{r},t)\mathbf{v}_\mathrm{g}d^3r.
 \label{eq:mdwmomentum}
\end{equation}
Here $\mathbf{v}_\mathrm{g}$ is the group velocity vector of the MP.
In numerical simulations described in Sec.~\ref{sec:simulations},
it is verified that both forms in Eq.~\eqref{eq:mdwmomentum} give
an equal result for light pulses in the narrow-band limit.

\subsection{Comparison of the OCD and MP quasiparticle models}
\label{sec:comparison}

For a light pulse with total electromagnetic energy $U_0$,
the total mass transferred by the MDW, given in Eq.~\eqref{eq:mdwmass2}, can also be written as
\begin{equation}
 \delta M=\int \rho_\text{\tiny MDW}d^3r
 \approx(n_\mathrm{p}n_\mathrm{g}-1)U_0/c^2.
 \label{eq:dM}
\end{equation}
The right hand side of Eq.~\eqref{eq:dM} equals the result obtained from the MP quasiparticle model \cite{Partanen2017e}.
Here the equality becomes accurate in the narrow-band limit.
For a narrow-band light pulse with central frequency $\omega_0$ in a lossless dispersive medium,
the total energy is given by
\begin{equation}
 E_\text{\tiny MP} =\int\Big\{\rho_\text{\tiny MDW}c^2+\frac{1}{2}\Big[\frac{d(\varepsilon\omega_0)}{d\omega_0}\mathbf{E}^2+\frac{d(\mu\omega_0)}{d\omega_0}\mathbf{H}^2\Big]\Big\}d^3r
 =\delta Mc^2+U_0\approx n_\mathrm{p}n_\mathrm{g}U_0.
 \label{eq:mpenergy}
\end{equation}
Here the first term is the mass energy of the MDW and the second
term is the conventional electromagnetic energy in the
electrodynamics of continuous media \cite{Landau1984}.
On the right, we have the total energy of the coupled MP state obtained
from the MP quasiparticle model \cite{Partanen2017e}.
The total momentum of the coupled MP state of the field and matter
can be written as
\begin{equation}
 \mathbf{p}_\text{\tiny MP}=\int\Big(\rho_\mathrm{a}\mathbf{v}_\mathrm{a}+\frac{\mathbf{E}\times\mathbf{H}}{c^2}\Big)d^3r
 \approx\frac{n_\mathrm{p}U_0}{c}\hat{\mathbf{x}}.
\end{equation}
The first term on the left is the MDW momentum in Eq.~\eqref{eq:mdwmomentum} and
the second term is the momentum density of the electromagnetic field.
On the right, we have the momentum density of the coupled MP state obtained
from the MP quasiparticle model \cite{Partanen2017e}. In the simulations
described in Sec.~\ref{sec:simulations}, it is found that
the OCD and MP quasiparticle model results agree in the narrow-band limit,
where the photon picture becomes reasonable.

\section{Simulations}
\label{sec:simulations}

\subsection{Visualization of the node structure of the MDW}
\label{sec:simulationsvisualization}

\begin{figure}
\centering
\includegraphics[width=\textwidth]{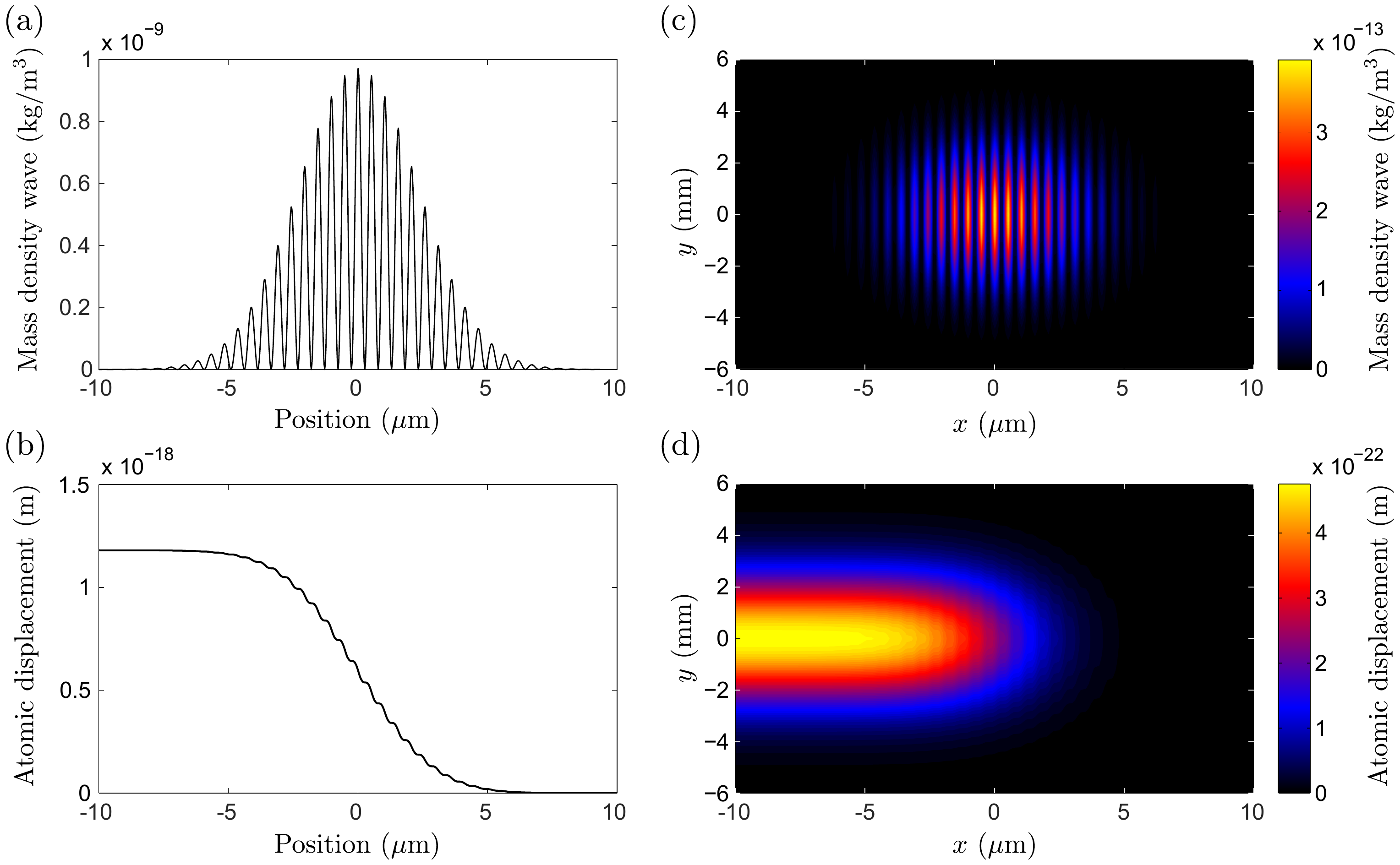}
\caption{\label{fig:linear}
(Color online) Simulation results for a
Gaussian light pulse having electromagnetic energy
$U_0=1$ $\mu$J, vacuum wavelength $\lambda_0=1550$ nm,
and $\Delta t_\mathrm{FWHM}=27$ fs. Panel (a) shows the MDW
and (b) shows the atomic displacements of a one-dimensional pulse
with energy $U_0$ per cross-sectional area of diameter $d=100$ $\mu$m.
The phase and group refractive indices of the material for the central frequency
are $n_\mathrm{p}=1.5$ and $n_\mathrm{g}=2$.
These parameters are close to technological feasibility limit
but they are suitable for our needs to visualize the node structure of the MDW.
Panel (c) shows the atomic mass density of the MDW and (d) shows
the atomic displacements of the MDW
in the plane $z=0$ m for a three-dimensional light
pulse having finite lateral dimensions.}
\vspace{-0.1cm}
\end{figure}

Next, we use the OCD model to simulate the propagation of one- and
three-dimensional Gaussian light pulses in a linearly dispersive material.
These simulations aim at illustrating the node structure of the
MDW and the related actual atomic displacements.
The field of a Gaussian light pulse is modeled using Eq.~\eqref{eq:electricfield}.
For the pulse, we use a vacuum wavelength of $\lambda_0=1550$ nm and
a total electromagnetic energy of $U_0=1$ $\mu$J.
In our example, the relative spectral width
of the pulse is $\Delta\omega/\omega_0=\Delta k_0/k_0=0.05$,
which corresponds to the temporal FWHM of $\Delta t_\mathrm{FWHM}=27$ fs.
This small value is close to the feasibility limit and it is used here
to make the node structure of the MDW visible in the same
length scale with the pulse envelope.
The spatial discretization used in the simulations is
$h_x=\lambda/40$, where $\lambda=\lambda_0/n_\mathrm{p}$ is
the wavelength in the medium, and the temporal discretization is $h_t=2\pi/(40\omega_0)$.
This discretization is sufficiently dense when compared
to the scale of the harmonic cycle.

In the simulations of this subsection, we use an artificial example material
that is suitable for our visualization needs.
The refractive indices of the example material
at the central frequency $\omega_0=2\pi c/\lambda_0$
are given by $n_\mathrm{p}=1.5$ and $n_\mathrm{g}=2$.
The value of the phase refractive index is close to typical values
for glasses but the group refractive index has been made
larger for visualization needs.
In the simulations, we assume a circular cross-sectional area $A=\pi(d/2)^2$
of diameter $d=100$ $\mu$m. This cross section is large enough so that the
resulting maximum value $3.3\times10^{11}$ W/cm$^2$
of the irradiance averaged over the harmonic cycle does not exceed the
bulk value of the breakdown threshold irradiance of many common materials, e.g.,
the value of $5.0\times10^{11}$ reported for fused silica \cite{Smith2007}.
For the equilibrium mass density of the material,
we use a value of $\rho_0=2400$ kg/m$^3$. The material is also assumed
to be isotropic having a bulk modulus of $B=50$ GPa and a shear modulus of $G=25$ GPa.
These material parameter values are close to typical values of
the corresponding quantities for glass.

The position dependence of the MDW obtained as
a result of the simulation is depicted in Fig.~\ref{fig:linear}(a)
when the center of the Gaussian light pulse is propagating to right at the position $x=0$ $\mu$m.
The MDW driven by the optoelastic forces follows from solving Newton's equation of motion
in Eq.~\eqref{eq:mediumnewton} and it equals the difference
$\rho_\text{\tiny MDW}(\mathbf{r},t)=\rho_\mathrm{a}(\mathbf{r},t)-\rho_0$
of the perturbed mass density $\rho_\mathrm{a}(\mathbf{r},t)$
and the equilibrium mass density $\rho_0$. The nodes and the
envelope of the MDW both follow the shape of the Gaussian
pulse as expected. The total transferred mass of the light pulse
can be obtained by the volume integral of the MDW mass density
in Fig.~\ref{fig:linear}(a). The resulting
numerical value of the transferred mass is $2.23\times 10^{-23}$ kg.
When this value is divided by the photon number of the pulse,
$N_0=U_0/\hbar\omega_0=7.8\times 10^{12}$,
we then obtain the value of
$1.60$ eV/$c^2$ for the transferred mass per photon.
This equals the MP quasiparticle value on the right
in Eq.~\eqref{eq:dM} within the relative error of $10^{-4}$.

The position dependence of the atomic displacements is shown in Fig.~\ref{fig:linear}(b).
These atomic displacements correspond to the MDW in Fig.~\ref{fig:linear}(a).
On the right of $x=7$ $\mu$m, the atomic displacement is zero since
the leading edge of the pulse propagating to right at $x=7$ $\mu$m has not yet reached this area.
Behind the pulse on the left of $x=-7$ $\mu$m, the atomic displacement has obtained a nonzero constant value of $r_\mathrm{a,max}=1.18\times10^{-18}$ m.
\linebreak
This is equal to $r_\mathrm{a,max}=\delta M/(\rho_0 A)$, where $\delta M$
is the total transferred mass of the light pulse in Eq.~\eqref{eq:dM}.
As the atoms behind the pulse have been shifted forwards and the atoms
on the right of the pulse are still at their original equilibrium positions,
the atoms are more densely spaced at the position of the light pulse.
This local increase in the density of atoms is the origin of the MDW in Fig.~\ref{fig:linear}(a).
At an arbitrary time, the momentum of atoms in the MDW is obtained as
an integral of the classical momentum density of atoms given in Eq.~\eqref{eq:mdwmomentum}.

We also simulate the MDW and the atomic displacements resulting from
the propagation of a three-dimensional Gaussian light pulse.
This light pulse is obtained from the one-dimensional pulse
in Eq.~\eqref{eq:electricfield} by adding
additional $y$ and $z$ dependencies using factors
$e^{-(\Delta k_y)^2y^2/2}$ and $e^{-(\Delta k_z)^2z^2/2}$.
Therefore, this light pulse is only an approximative solution of
Maxwell's equations. However, as reasoned in Ref.~\cite{Partanen2017c},
this approximation becomes accurate in the limit where
$\Delta k_y$ and $\Delta k_z$ are small compared to the wave number
in the medium, given by $k_\mathrm{0,med}=n_\mathrm{p}k_0$.
Here we use relatively small values $\Delta k_y=\Delta k_z=10^{-4}k_0$,
which make this approximation sufficiently accurate for our visualization purposes.

The MDW of the three-dimensional Gaussian light pulse
is illustrated in Fig.~\ref{fig:linear}(c) by a contour
plot in the plane $z=0$ m. The main difference between
the three- and one-dimensional light pulses
is formed by the finite lateral dimensions of the three-dimensional
pulse as described above. This can also be seen in the MDW in Fig.~\ref{fig:linear}(c),
where the atomic density of the MDW becomes zero for increasing and decreasing values of $y$.
From the lateral dimensions used and the corresponding
smaller value of the energy per cross-sectional area,
it also follows that the values of the atomic density of
the MDW in Fig.~\ref{fig:linear}(c) are smaller than
the corresponding values of the one-dimensional MDW in Fig.~\ref{fig:linear}(a)

Figure \ref{fig:linear}(d) shows the contour plot of the $x$ component
of the atomic displacements resulting from the propagation of the three-dimensional
Gaussian pulse in the plane $z=0$ m. These atomic displacements correspond
to the MDW in Fig.~\ref{fig:linear}(c). As a result of the smaller value of the
energy per cross-sectional area, the values of the atomic displacement in Fig.~\ref{fig:linear}(d)
are also smaller than the values in the one-dimensional case in Fig.~\ref{fig:linear}(b).

\subsection{Estimating atomic displacements of the MDW in silicon}

\begin{figure}
\centering
\includegraphics[width=\columnwidth]{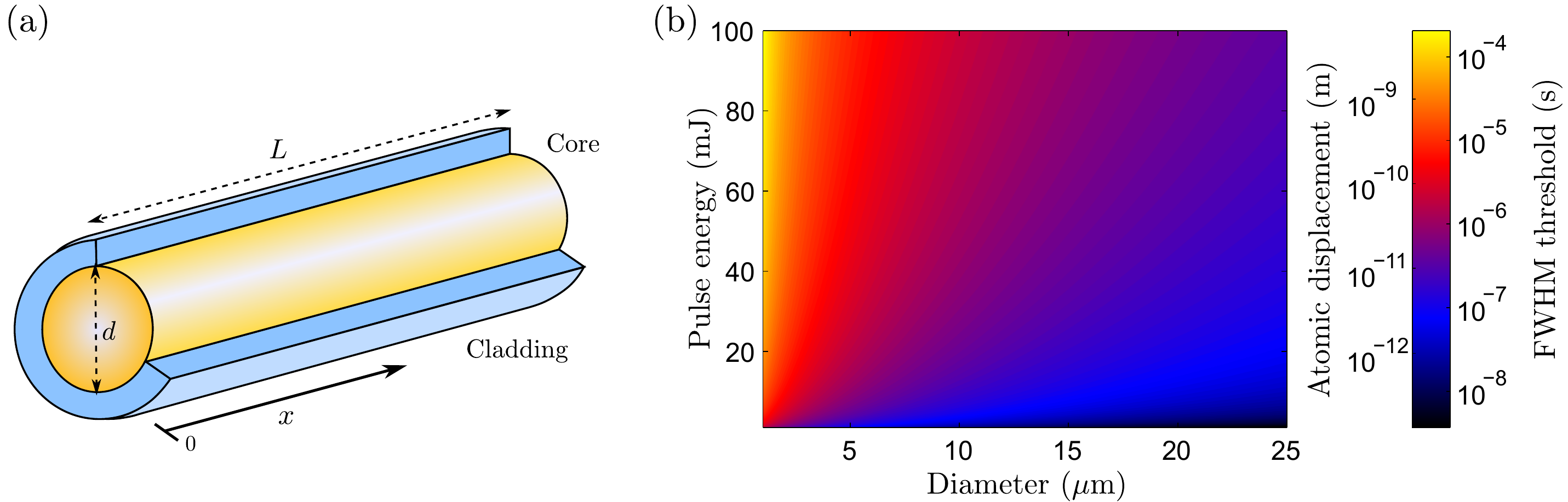}
\caption{\label{fig:fibervariation}
(Color online) (a) Schematic illustration of an optical fiber
with a core diameter $d$ and length $L$. The transferred mass of the MDW
can be calculated by measuring the shift of atoms on the surface of the waveguide
at $x=L/2$ just after the light pulse has gone.
(b) The dependence of the total atomic displacement of the MDW of a Gaussian
light pulse in silicon on the pulse energy and on the diameter of
the cross-sectional area of the pulse.
The vacuum wavelength is $\lambda_0=1550$ nm and the corresponding
phase and group refractive indices are $n_\mathrm{p}=3.4757$ and $n_\mathrm{g}=3.5997$. 
The second color-bar axis shows the threshold
$\Delta t_\mathrm{th}$ of the pulse. It is  obtained by requiring
that the energy density of the pulse does not exceed the bulk value
of the breakdown threshold irradiance.}
\end{figure}

Next, we study the dependence of the atomic displacement of the MDW
on the pulse energy and on the diameter of the pulse cross section. These
simulations serve for designing experimental setups for the measurement of the
transferred mass of a light pulse. We assume a one-dimensional Gaussian pulse
in silicon and calculate the MDW shift for selected pulse energies,
cross-sectional areas, and $\Delta t_\mathrm{FWHM}$.
The simulated atomic displacements correspond to the experimental setup in which the
pulse energy is propagating in a waveguide or an optical fiber as depicted in Fig.~\ref{fig:fibervariation}(a).
Due to the interface effects, the physical cross-sectional area of the fiber is not equal
to the cross-sectional area of our calculations. The effective core cross section of the fiber
must account for the possible cladding layer, metallic coating, and other factors that
govern the spreading of the pulse energy in the transverse direction. In quantitative
predictive simulations, the waveguide dispersion should also be taken into account.
All these factors can be included in the OCD simulations.

For the phase and group refractive indices of crystalline silicon,
we use literature values $n_\mathrm{p}=3.4757$ and 
$n_\mathrm{g}=3.5997$ respectively for $\lambda_0=1550$ nm \cite{Li1980}. For the density, we use
$\rho_0=2329$ kg/m$^3$ \cite{Lide2004}. The elastic
constants in the direction of the (100) plane used in the simulations
were $C_{11}=165.7$ GPa, $C_{12}=63.9$ GPa, and $C_{44}=79.6$ GPa
\cite{Hopcroft2010}. These elastic constants give for the
bulk modulus a value of $B=(C_{11}+2C_{12})/3=97.8$ GPa and
for the shear modulus a value of $G=C_{44}=79.6$ GPa.

Figure \ref{fig:fibervariation}(b) depicts the dependence
of the atomic displacement on the pulse energy and on the diameter of the
cross-sectional area. Here we assume much longer pulses than above,
typically $\Delta t_\mathrm{FWHM}>1$ ns.  Therefore, the correspondence between the
MP and the OCD models is very accurate. Thus we use the quasiparticle model result
$r_\mathrm{a,max}=\delta M/(\rho_0A)$ for the maximum atomic displacement $r_\mathrm{a,max}$.
We use  $\delta M=(n_\mathrm{p}n_\mathrm{g}-1)U_0/c^2$
and $A=\pi(d/2)^2$, where $d$ is the diameter of the cross-sectional area,
to obtain $r_\mathrm{a,max}=(n_\mathrm{p}n_\mathrm{g}-1)U_0/[c^2\rho_0\pi(d/2)^2]$.
The atomic displacement depends linearly on the
pulse energy and it is inversely proportional to
the cross-sectional area.
Accordingly, one can see in Fig.~\ref{fig:fibervariation}(b) that
the atomic displacement is large for high pulse energies
and for small cross-sectional areas.

\subsubsection{Influence of the material breakdown irradiance}

We also represent in Fig.~\ref{fig:fibervariation}(b)
the minimum $\Delta t_\mathrm{FWHM}$ of a Gaussian pulse
that is needed to produce the corresponding atomic displacement without
exceeding the bulk value of the breakdown threshold
irradiance of the pertinent material. Using the total electromagnetic pulse energy denoted by $U_0$,
and the cross-sectional area of the pulse 
$A=\pi(d/2)^2$, this threshold
$\Delta t_\mathrm{FWHM}$, denoted by $\Delta t_\mathrm{th}$, is calculated as 
$\Delta t_\mathrm{th}=2U_0/[\pi(d/2)^2I_\mathrm{th}]$. Here
$I_\mathrm{th}$ is the bulk value of the breakdown threshold
irradiance of the material. The corresponding value of the
fluence is given by $F_\mathrm{th}=2U_0/[\pi(d/2)^2]$.
The factor $2$ comes from the fact that
the pulse is Gaussian instead of a top-hat pulse with
constant irradiance. For silicon with $\lambda_0=1550$ nm,
the bulk value of the breakdown threshold energy
density $u_\mathrm{th}=13.3$ J/cm$^3$ has been reported by Cowan \cite{Cowan2006}.
This corresponds to the threshold irradiance of
$I_\mathrm{th}=u_\mathrm{th}v_\mathrm{g}=1.11\times 10^{11}$ W/cm$^2$.
These are values averaged over the harmonic cycle.

The calculated threshold $\Delta t_\mathrm{th}$ of a Gaussian pulse is presented by the second
color-bar axis in Fig.~\ref{fig:fibervariation}(b).
Using the formulas given above, the scaling between
the atomic displacement and the threshold $\Delta t_\mathrm{th}$
is given by $r_\mathrm{a,max}/\Delta t_\mathrm{th}=(n_\mathrm{p}n_\mathrm{g}-1)I_\mathrm{th}/(2c^2\rho_0)$.
This indicates that, to obtain
large atomic displacements for a given pulse
energy, one should choose a material
with a high refractive index, high breakdown
threshold irradiance, and relatively small mass
density. Fig.~\ref{fig:fibervariation}(b) shows that,
in order to obtain atomic displacements larger than $1$ nm
\linebreak
in silicon without breaking the material, a pulse width 
larger than $\Delta t_\mathrm{th}=33$ $\mu$s must be used.

\subsubsection{Displacement of atoms due to optical absorption}

In the simulations above, we have neglected
possible momentum transfer and shift of atoms caused by light absorption.
To estimate the accuracy of this approximation, we next calculate the total atomic displacement
and the atomic velocity caused by the optical absorption.
The mass of a cylindrical medium block with a diameter $d$ and length $L$,
i.e., the core of the fiber in Fig.~\ref{fig:fibervariation}(a),
is given by $M=\rho_0\pi(d/2)^2L$.
The momentum absorbed by the cylinder is given by
$P_\mathrm{abs}=(1-e^{-\alpha L})n_\mathrm{p}U_0/c\approx \alpha L n_\mathrm{p}U_0/c$,
where $\alpha$ is the small absorption coefficient of the medium.
The velocity of the medium block resulting from the absorption is then
$V_\mathrm{abs}=P_\mathrm{abs}/M\approx \alpha n_\mathrm{p}U_0/[c\rho_0\pi(d/2)^2]$.
In the time scale of $\Delta t_\mathrm{FWHM}$, this gives
an atomic displacement of
$X_\mathrm{abs}=V_\mathrm{abs}\Delta t_\mathrm{FWHM}$.

For single-crystal silicon, absorption is very low at $\lambda_0=1550$ nm.
Schinke \emph{et al.}~\cite{Schinke2015} and Green \cite{Green2008} have given
for $\lambda_0=1450$ nm give $\alpha\approx10^{-8}$ cm$^{-1}$ and the absorption
decreases towards $\lambda_0=1550$ nm. Thus, we conservatively estimate
$\alpha=10^{-8}$ cm$^{-1}$. We use $\Delta t_\mathrm{FWHM}=\Delta t_\mathrm{th}=33$ $\mu$s and
$d=2.5$ $\mu$m corresponding to $r_\mathrm{a,max}=1.0$ nm atomic
displacement due to the MDW. By solving the threshold pulse energy
from $\Delta t_\mathrm{th}=2U_0/[\pi(d/2)^2I_\mathrm{th}]$,
we then obtain $U_0=90$ mJ.
The corresponding velocity of atoms is $V_\mathrm{abs}=9.1\times 10^{-8}$ m/s and,
in the time intervall of $\Delta t_\mathrm{th}$, the resulting
atomic displacement is given by
$X_\mathrm{abs}=3.0$ pm. Thus the atomic displacement due to optical absorption
is vastly smaller than $r_\mathrm{a,max}=1.0$ nm
following from the MDW. We conclude that
optical absorption should not prevent
measurements of the atomic displacements due to the MDW. 
The experimental measurement of the transferred mass
of the MDW is feasible in the light of these results.

\subsubsection{Thermal expansion following from optical absorption}
Next, we consider thermal expansion
following from the optical absorption.
If the rate of absorption of optical energy
remains low compared to the total field energy,
as is the case in our example, the absorbed energy
per unit length is approximately constant. This energy will be
converted into heat and it leads to a local equilibrium lattice and electronic
temperature in a relaxation time that is of the order of picoseconds.
This elevated equilibrium temperature then corresponds to an
increased lattice constant and gives rise to an elastic force
that starts to shift atoms towards their new equilibrium positions.
Thus, the thermal expansion takes place at the velocity
of sound and comes behind the optical pulse and the
MDW shift of atoms which is a shock wave propagating
at the velocity of light.
Using the literature values for the specific-heat capacity and
thermal-expansion coefficients, it can be estimated that
the thermal expansion does not lead to measurable atomic
displacements in the middle part of the fiber in a
time $\Delta t_\mathrm{FWHM}$ which is shorter than the
time the sound waves need to travel through the fiber.
Thermal expansion is also related to the longitudinal relaxation studied
below.

\subsubsection{Transverse relaxation}

In measuring the transferred mass of the MDW, one also has to account for
the relaxation of the atomic displacements of the MDW by phonons. 
This relaxation  is governed by the elastic forces in the OCD model and
it takes place at the velocity of sound.

The relaxation effect has been represented in Ref.~\cite{Partanen2017c}.
If a three-dimensional light pulse propagates in the core of an optical
fiber that has a cladding, the MDW displaces atoms as shown
in Fig.~\ref{fig:linear}(d). Along the path of the MP, the atoms are 
displaced forwards. The atoms in the surrounding layers are
not shifted forwards since they do not feel the optical force.
This gives rise to a shear strain field along the path of the MP.
By the transverse relaxation we mean in the following the relaxation of the strain field
so that atoms in the displaced region (core) are shifted backwards and atoms
in the surrounding layers (cladding) are shifted forwards. After this relaxation,
the longitudinal strain becomes constant across the cross section
of the waveguide.

In optical fibers, the transverse relaxation of the strain field is fast
since the distances to be traveled by phonons in the transverse direction are very short.
The longitudinal velocity of sound in silicon is
given by $v_\parallel=\sqrt{C_{11}/\rho_0}=8435$ m/s. The time that it takes for a sound
wave to propagate, e.g., a distance of $1$ mm is $1.2$ ns. This is a short time when
compared to the time scale of $\Delta t_\mathrm{FWHM}=33$ $\mu$s used above.
Thus, the transverse relaxation takes place in a time
scale that is shorter than the passing time of the pulse.
Atoms are accordingly displaced in the longitudinal direction
by the same amount in the middle and at the surface of the waveguide. When the transverse
relaxation has taken place, the atomic displacement is constant in the transverse plane and it is reduced to
$r_\mathrm{a,relaxed}=r_\mathrm{a,max}\rho_0A/(\rho_\mathrm{eff}A_\mathrm{tot})$.
Here $A_\mathrm{tot}$ is the total cross-sectional area of all layers in the given transverse plane and
$\rho_\mathrm{eff}$ is the effective mass density of the cross-sectional
area given by $\rho_\mathrm{eff}=\sum_i\rho_iA_i/A_\mathrm{tot}$.
Here the sum is taken over all material layers. The densities and
cross-sectional areas of the corresponding layers are denoted by $\rho_i$ and $A_i$.

In summary, the time needed for the transverse relaxation is much
shorter than the pulse width in the time domain $\Delta t_\mathrm{FWHM}$ for
structures where the atomic displacement is potentially measurable.
It is thus advantageous to keep the waveguide diameter as small
as possible. The tradeoff is to be made by considering the effectivity of the coupling
of the light source to the waveguide and the technical
processing aspects of fabricating the waveguide.
Within these engineering feasibility limits, the narrower
the waveguide is, the larger is the atomic displacement and
the breakdown of the material can be prevented by
increasing the pulse width $\Delta t_\mathrm{FWHM}$.
There is however a further limiting factor that will set
a limit for increasing $\Delta t_\mathrm{FWHM}$,
based on the longitudinal relaxation.
This will be described in the next subsection.

\subsubsection{Longitudinal relaxation}

After the transverse relaxation is complete, further relaxation
starts by sound waves from the ends of the fiber.
If the fiber is long enough, the relaxation does not have
time to reach the middle part of the fiber where
the atomic displacement is to be measured. In the time scale of $\Delta t_\mathrm{FWHM}=33$ $\mu$s,
the distance traveled by sound in silicon 
is $28$ cm. A fiber with length $L>56$ cm is thus enough
to avoid the longitudinal relaxation from influencing the measured
atomic displacement if the displacement in
the middle part of the fiber is measured
immediately after the light pulse has passed this zone.

\section{Conclusions}
\label{sec:conclusions}

In conclusion, we have generalized the recently developed MP
theory of light for the study of light propagation in dispersive
media. We have compared the related OCD model to the MP quasiparticle picture. Our results
show how the total energy and momentum are shared by the electromagnetic field
and the medium atoms in the MDW. Both the MP quasiparticle picture and the OCD approach lead to transfer
of mass with the light pulse. The transfer of mass comes from the shift
of atoms in the medium and it is expected to be experimentally feasible to measure. 
To facilitate the planning of measurements, we
have carried out simulations of how the atomic displacements
due to the MDW changes as a function of space and time in a simple silicon
waveguide structure. In the simulations, we have considered different
waveguide cross sections and optical pulse widths. We have also accounted
for the breakdown threshold irradiance of silicon,
which is one of the main limiting factors in possible experimental setups
as the electromagnetic energy density in the medium cannot be made
arbitrarily large without damaging the medium. Using the OCD model,
it is also possible to perform more detailed simulations, which account
for the waveguide dispersion and the spreading of the pulse energy
in the transverse direction.
These simulations are a topic of further work.

\begin{acknowledgments}
This work has in part been funded by the Academy of Finland and the Aalto Energy Efficiency Research Programme.
\end{acknowledgments}


\end{document}